\documentclass[12pt]{article}
\usepackage{graphicx,cite,amssymb,epsfig,float,psfrag,multirow,amsmath,
subfigure}
\catcode`@=11

\def\@citex[#1]#2{\if@filesw\immediate\write\@auxout{\string\citation{#2}}\fi
  \def\@citea{}\@cite{\@for\@citeb:=#2\do
    {\@citea\def\@citea{,\penalty\@m}\@ifundefined
      {b@\@citeb}{{\bf ?}\@warning
       {Citation `\@citeb' on page \thepage \space undefined}}%
\hbox{\csname b@\@citeb\endcsname}}}{#1}}

\def\citer{\@ifnextchar [{\@tempswatrue\@citexr}{\@tempswafalse\@citexr[]}}

\def\@citexr[#1]#2{\if@filesw\immediate\write\@auxout{\string\citation{#2}}\fi
  \def\@citea{}\@cite{\@for\@citeb:=#2\do
    {\@citea\def\@citea{--\penalty\@m}\@ifundefined
       {b@\@citeb}{{\bf ?}\@warning
       {Citation `\@citeb' on page \thepage \space undefined}}%
\hbox{\csname b@\@citeb\endcsname}}}{#1}}

\def\citer{\@ifnextchar [{\@tempswatrue\@citexr}{\@tempswafalse\@citexr[]}}
\catcode`@=12

\newcommand{\beq}{\begin{eqnarray}}
\newcommand{\eeq}{\end{eqnarray}}

\newcommand{\nn}{\nonumber}
\newcommand{\lsim}{\stackrel{<}{_\sim}}

\def\gsim{ {\ \lower-1.2pt\vbox{\hbox{\rlap{$>$}\lower5pt\vbox{\hbox{$\sim$}}}}\ } }
\begin{document}

\begin{flushright}
{\tt CERN-PH-TH/2004-221}\\
{\tt LMU 14/04}\\
{\tt November 2004}
\end{flushright}

\vspace*{1.5cm}
{\centerline{\Large \bf Predictions for $B \to K \gamma \gamma$ decays}}
\vspace*{2.cm}
\centerline{
{\sc Gudrun Hiller$^{1,2}$\footnote{
Email address: 
hiller@theorie.physik.uni-muenchen.de} and  A.~Salim~ Safir$^2$\footnote{Email address: safir@theorie.physik.uni-muenchen.de}
}
}
\smallskip
\begin{center}\sl $^1$ CERN, TH Division, Dept.\ of Physics, 
CH-1211 Geneva 23, Switzerland\\[.3em]

$^2$ Ludwig-Maximilians-Universit\"at M\"unchen, 
Department f\"ur Physik, \\
Theresienstra\ss e 37, D-80333 Munich, Germany
\end{center}

\vspace*{1.5cm}

\begin{abstract}
We present a phenomenological study of the rare double radiative decay 
$B\to K \gamma\gamma$ in the Standard Model (SM) and beyond. 
Using the operator product expansion (OPE) technique, we estimate the
short-distance (SD) contribution to the decay amplitude in a region of 
the phase space which is around the point where all decay products have 
energy $\sim m_b/3$ in the rest frame of the $B$-meson.
At lowest order in $1/Q$, where $Q$ is of order $m_b$, 
the $B\to K \gamma\gamma$ matrix element is then 
expressed in terms of the usual  $B\to K$ form factors known from
semileptonic rare decays. 
The integrated SD branching ratio in the SM in the OPE region turns out
to be $\Delta {\cal{B}}(B \to K \gamma \gamma)_{SM}^{OPE} \simeq 1 
\times 10^{-9}$.
We work out the di-photon invariant mass 
distribution with and without the resonant  background through
$B\to K \{\eta_c,\chi_{c0}\}\to K\gamma \gamma$.
In the SM, the resonance contribution is dominant in the region of phase space 
where the OPE is valid.
The present experimental upper limit on $B_s \to \tau^+ \tau^-$ 
decays, which constrains the scalar/pseudoscalar 
Four-Fermi operators with $\tau^+ \tau^-$, leaves considerable room for 
new physics in the one-particle-irreducible contribution
to $B\to K \gamma \gamma$ decays. 
In this case, we find that the SD $B\to K \gamma \gamma$ branching ratio
can be enhanced by one  order of magnitude with respect to its SM value
and the SD contribution can lie outside of the resonance peaks.

\end{abstract}

\vspace{1cm}
\noindent

\newpage
\section{Introduction}
Rare $B$-decays mediated by flavor changing 
neutral currents (FCNC)
provide a rich laboratory to study
the flavor structure of the Standard Model (SM) and its extensions. 
Such processes are forbidden in the Born approximation 
and are generated by 
loops, which gives them a high sensitivity to new physics (NP). 
A prominent example is the both experimentally and theoretically 
well-studied $b\to s\gamma$ transition, 
exemplified by the inclusive $B\to X_s \gamma$
and the exclusive $B \to K^{*}\gamma$ decays 
(for a detailed review, see \cite{Ali-rev}). 

Radiative $B$-decays involving two photons, such as 
$B \to\gamma\gamma$, $B\to X \gamma\gamma$ 
or $B\to K^{(*)} \gamma\gamma$ decays, are of further interest. 
Among the Cabibbo-favored $b \to s $ transitions,
the decays $B_s \to \gamma \gamma$ and  
$B \to X_s  \gamma \gamma$  have been theoretically 
analyzed in the SM some 
time ago~\cite{Lin:1989vj} and later on by taking into account 
the leading-order QCD corrections in the framework of the electroweak 
effective Hamiltonian~\citer{Hiller:1997ie,Bosch:2002bv}.
Recalling that at the quark level the FCNC double-photon decay is mediated 
by $b \to s \gamma \gamma$, it receives one-particle-reducible (1PR) 
contributions from the $b \to s \gamma $ transition plus an additional 
photon and a one-particle-irreducible (1PI) term from a fermion loop with 
the two photons emitted from that loop.
The correlation with $b \to s \gamma$ implies that NP, which 
predominantly induces
contributions to the 1PR piece, gives no drastic effects
in the double radiative decays
because of the strong constraints from  $B \to X_s \gamma$ data.
This has been shown, for example, for
the two-Higgs-doublet model~\cite{Reina:1996up}.
On the other hand, NP in the 1PI contribution, that is in Four-Fermi operators,
is subleading in  $b \to s \gamma$ decays and hence
can be sizable in $b \to s \gamma \gamma$ transitions where it enters at the
same order as the 1PR contributions.
So far only experimental upper limits on double radiative decays 
have been set (at $90 \%$ C.L.):
\begin{eqnarray}
&&{\cal{B}}(B_d \to \gamma \gamma)  < 1.7 \times 10^{-6}
~~\mbox{(BaBar) \cite{Aubert:2001fm}} \, , \nn\\
&&{\cal{B}}(B_s \to \gamma \gamma) < 1.48 \times 10^{-4}
 ~~\mbox{(L3) \cite{Acciarri:1995hv}},
\end{eqnarray}
which are about two orders of magnitude above their respective
SM values, e.g.~\cite{Bosch:2002bv}.

While many theoretical investigations have been dedicated to 
$B_s \to \gamma \gamma$ and  $B \to X_s  \gamma \gamma$,
less attention has been devoted to exclusive 
$B \to K^{(*)} \gamma \gamma$ decays.
For instance, only two groups~\cite{Singer:1997ti,Choudhury:2002yu} 
have estimated the $B \to K \gamma \gamma$ branching ratio, with the 
result given as $\sim (0.5 -5.6)\times 10^{-7}$, 
depending on cuts on the photon energies.
Note that $B \to K \gamma \gamma$ decays have a complicated matrix element
due to the non-local 1PR contribution. In the aforementioned 
studies the 1PR piece
has been solely parametrized in terms of a long-distance contribution
using hadronic models, i.e.~vector meson dominance.
There is also a sizable resonant background from charmonia through
$B\to K (c \bar c) \to K\gamma \gamma$. 
Since $B \to K \gamma \gamma$ decays provide a reasonable experimental signal 
for not too soft photons and 
with branching ratios in the $10^{-7}$ range, they are in 
reach of the currently operating $B$-factories at SLAC and KEK and at
experiments at the hadron colliders, Tevatron and LHC.
Therefore, a detailed study of these decays in the SM and beyond is of great 
interest, in particular due to the complementarity with $b \to s \gamma$ decays
with respect to NP searches.

Our main objectives in this paper are {\it i}
to write down a model-independent description of 
exclusive $B \to K \gamma \gamma$ decays
and {\it ii} to find out to what extend one can extract short-distance (SD)
physics from these decays, that is, how big is the SD contribution 
and can it be experimentally isolated from the resonance contributions.
The idea here is to use the freedom of a 3-body decay
and choose the photon energies such that the
virtualities of the intermediate quarks in the 1PR $b \to s \gamma \gamma$ 
diagrams are hard. Then, these scales of order $m_b$ are integrated out with
the effective field theory methods using heavy-quark-effective-theory 
(HQET)~\cite{Neubert:1993mb} and the 
soft-collinear-effective-theory (SCET) \cite{Bauer:2000yr,Beneke:2002ph}.
The resulting operator product expansion 
(OPE) enables us to express the matrix element in terms of the 
$B \to K$ form factors in some region of the phase space. 
There are other mechanisms contributing to $B \to K \gamma \gamma$ decays, 
such as when one photon gets emitted from the 
spectator quark or annihilation diagrams, which we comment on.
We further incorporate the $\eta_c,\chi_{c0}$-resonance contributions 
using a phenomenological parametrization \`a la Breit-Wigner to see whether
they can be distinguished from the SD ones obtained from the OPE.
We work out possible NP effects in the $B \to K \gamma \gamma$ spectra
and branching ratios. In particular, we illustrate the impact of a  
scenario with enlarged 
$(\bar s b) (\tau^+ \tau^-$) couplings, which
substantially modifies the 1PI $b \to s \gamma \gamma$ contribution
 but does not violate existing constraints on any other rare $B$-decays.

The remainder of this paper is organized as follows: After a brief description 
of the effective Hamiltonian formalism for double radiative 
$b \to s \gamma \gamma$ decays in Section \ref{sec:bsgg}, we evaluate  
in Section \ref{sec:BKgg} 
the (non-local) matrix element of $B \to K \gamma \gamma$ decays
with the OPE and present the resulting decay amplitude and 
distributions.
Additional contributions to the $B\to K\gamma \gamma$ amplitude, namely  
photon radiation off the spectator, annihilation 
and  contributions from intermediate charmonia are 
explored in Section \ref{sec:addi}.
In Section \ref{sec:pheno}, 
a phenomenological discussion of the $B\to K\gamma \gamma$ Dalitz region, 
in which the OPE is valid,
and the di-photon invariant mass 
distributions is given in the SM and with NP.
We summarize in Section \ref{sec:conclusions}.

%
\section{The lowest order $b \to s \gamma \gamma$ amplitude \label{sec:bsgg}}
We treat the effects of heavy degrees of freedom at the weak scale
with an effective Hamiltonian known from $b \to s \gamma$ decays, 
e.g.~\cite{Buchalla:1995vs}
\beq
{\cal{ H}}_{\rm{eff}}&=&-\frac{4 G_F}{\sqrt{2}} V_{tb} V_{ts}^* 
\sum_{i}
 C_i(\mu) {\cal{O}}_{i}(\mu), \label{eq:heff}
\eeq
which contains four-fermion operators
\begin{align}
{\cal{O}}_1&=(\bar{s}_\alpha \gamma^{\mu} L c_\beta)(\bar{c}_\beta
\gamma_{\mu} L b_\alpha), &
{\cal{O}}_2& =(\bar{s}_\alpha
\gamma^{\mu} L c_\alpha)(\bar{c}_\beta \gamma_{\mu} L b_\beta),
\label{eq:O12}
\end{align}
and the electromagnetic dipole operator
\begin{equation}
{\cal{O}}_{7}=\frac{e}{16 \pi^2} m_b \bar{s}_{\alpha} \sigma^{\mu \nu} R
b_{\alpha} F_{ \mu \nu}. \label{eq:O7}
\end{equation}
Here, $\alpha$ and $\beta$ are color indices, $L,R$=$(1  \mp\gamma_5)/2$ are chiral projectors and 
$F_{\mu \nu}$  denotes the QED field strength tensor.
In  Eq.~(\ref{eq:heff}) we neglect the penguin operators 
$\bar s \gamma_\mu L b \sum \bar q \gamma^\mu L/R q$ and double 
Cabibbo-suppressed contributions proportional to $V_{ub} V_{us}^* $.
We further neglect the strange quark mass throughout this paper.
Analytical formulas for the Wilson coefficients $C_{i}(\mu)$ 
can be seen in \cite{Buchalla:1995vs}.
Using the equations of motion, one can show that there are not further
independent, gauge invariant operators with two photons and with 
dimension equal or less
than six. Hence, the Hamiltonian in Eq.~(\ref{eq:heff}) is the same for 
$b \to s \gamma$ and $b \to s \gamma \gamma$ decays.

\begin{figure}[t]
\begin{center}
\epsfig{file=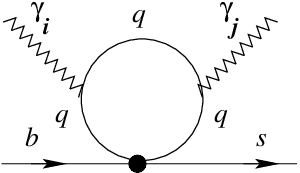,width=4cm,height=2.2cm}
\hspace*{2.cm}\epsfig{file=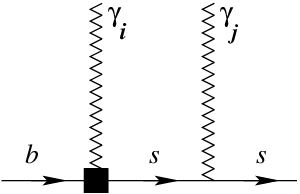,width=4cm,height=2.2cm}
\hspace*{0.5cm}\epsfig{file=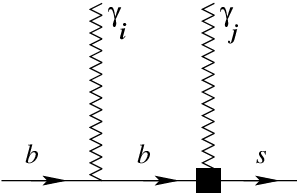,width=4cm,height=2.2cm}
\end{center}
\caption{\it Leading order Feynman diagrams 
contributing to $b\to s \gamma\gamma$ decays.
The 1PI diagram illustrates the insertion of four-fermion operators
(left plot), while the 1PR diagrams (right two plots) represent the 
insertion of $O_7$. 
The indices $i,j =1,2$ (with $i\neq j$) denote both photons and their 
interchanged counterparts.
\label{fig1}}
\end{figure}

The lowest order Feynman diagrams contributing to $b\to s \gamma\gamma$ 
decays are shown in Fig~\ref{fig1}. Only the sum of the two 1PR 
diagrams is gauge invariant \cite{Lin:1989vj}. The corresponding amplitude 
 can be written as \cite{Reina:1997my} 
\begin{eqnarray}
{\cal{A}}(b \to s \gamma \gamma)& =&
-\frac{i e^2 G_F}{\sqrt{2} \pi^2} V_{t b} V_{t s}^* 
\bar{u}_{s}(p^\prime)  \cdot \left [ (N_C C_1 +C_2) Q_u^2 \kappa_c W_2^{\mu \nu} 
\right. \\
&+& \left. C_7  Q_d  W_7^{\mu \nu} \right ] \cdot  u_b(p) 
\epsilon_{\mu}(k_1) \epsilon_{\nu}(k_2) \nonumber ,
\end{eqnarray}
where 
$p (p^\prime)$ represents the momentum of the $b(s)$-quark. We
denote by $k_1,k_2$ the 4-momenta and by $\epsilon(k_{1,2})$ 
the polarization vectors of the photons.
The auxiliary tensors  are written as
\beq
W_2^{\mu \nu} =\hspace*{-0.6cm}&&- i \left \{  \frac{1}{k_1 \cdot k_2} \left [ k_1^{\nu}
\epsilon^{\mu \rho \sigma \lambda} \gamma_{\rho}
k_{1 \sigma} k_{2 \lambda}
- k_2^{\mu} \epsilon^{\nu \rho \sigma \lambda} \gamma_{\rho}
 k_{1 \sigma} k_{2 \lambda} \right ]+ \epsilon^{\mu \nu \rho  \lambda}
  \gamma_{\rho} (k_2-k_1)_{\lambda}  \right \} L,~~~~\label{eq:W7}\\
W_7^{\mu \nu}=\hspace*{-0.6cm}&&\frac{m_b}{2} \left [
\frac{\not\! k_1 \gamma^{\mu} R ( \not\! p- \not\! k_2+m_b)
\gamma^{\nu}}{(p-k_2)^2-m_b^2} 
+\frac{\gamma^{\nu} (\not\! p^{\, \prime} + \not\!k_2)
\not\! k_1 \gamma^{\mu} R}
{(p^\prime+k_2)^2} \right]  
 + \left( k_1,\mu \leftrightarrow k_2 ,\nu \right),
\eeq
where $N_C=3$ denotes the number of colors, $Q_u=2/3$ and $ Q_d=-1/3 $
are  the up-type and down-type quark electric charges. 
Note that $W_2^{\mu \nu}$ 
is proportional to the Adler-Rosenberg tensor~\cite{AR}. We further suppress 
the renormalization scale dependence of the Wilson coefficients.
To the order we are working, the Wilson coefficients are evaluated at 
leading log at $\mu \simeq m_b$. We further use $\epsilon^{0123}=-1$.
The loop function $\kappa_q$ is
defined as \cite{ag91}
\beq \kappa_q &=& \kappa(q^2,m_q^2)=\frac{1}{2}+\frac{1}{z_q} \int_0^1
\frac{d x}{x} \ln \left (1-z_q x + z_q x^2 \right ),  \nonumber
\\ & =& \left \{  \begin{array}{ll}  \frac{1}{2}-\frac{2}{z_q}
\left (\arctan \sqrt{\frac{z_q}{4-z_q}}  \right )^2  &   {\rm if
}\ \  z_q<4~, \\ \frac{1}{2}+\frac{1}{z_q} \left
[-\frac{\pi^2}{2}+2 \left ( \ln \frac{\sqrt{z_q}+\sqrt{z_q-4}}{2}
\right ) ^2- 2 i  \pi \ln \left (\frac{\sqrt{z_q}+\sqrt{z_q-4}}{2}
\right ) \right ] \ \ \  & {\rm otherwise}, \end{array}   \right.
\label{eq:kappa}
\eeq
where $z_q = q^2/m_q^2$ and $q^2=(k_1+k_2)^2$.
It assumes the following limits
\begin{eqnarray} \label{eq:kappalimit}
\kappa_c =  \left \{  \begin{array}{ll} -\frac{q^2}{24 m_c^2} + {\cal{O}}(q^4/m_q ^4) & \mbox{for} ~~ q^2 \ll m_c^2,  \\ \frac{1}{2} + {\cal{O}}(m_c^2/q^2,m_c^2/q^2 \times logs) & \mbox{for} ~~ q^2 \gg m_c^2.  \end{array} \right.
\end{eqnarray}
%
\section{$B \to K \gamma \gamma$ decays \label{sec:BKgg} }
In this section we obtain explicit expressions for the 
$ B \to K \gamma \gamma$\footnote{Since we are not concerned about CP 
violation and to avoid clutter we use the convention $B \equiv b \bar q$ throughout this work.}
 decay distributions.
A general parametrization of the
amplitude and decay rate is given in Section \ref{sec:general}.
We perform an expansion in inverse scales of order
$m_b$ in Section \ref{sec:ope} to render the matrix element of the 1PR contribution local. This enables us to express the
$ B \to K \gamma \gamma$ matrix element  
in terms of $B \to K$ form factors. The resulting  
spectrum is presented in Section \ref{sec:spectrum}.

\subsection{The general amplitude and decay spectra \label{sec:general}}

We write the general $ B \to K \gamma \gamma$ amplitude as
\begin{equation}
{\cal {A}}(B \to K \gamma \gamma)=T_{\mu \nu} \epsilon^{ \mu}(k_1)
\epsilon^{ \nu}(k_2),
\end{equation}
where
\begin{eqnarray}
T_{\mu \nu } & =& A ( g_{\mu \nu} -\frac{k_{1 \nu} k_{2 \mu}}{k_1 \cdot
  k_2})
+i B \epsilon_{\mu \nu \alpha \beta} k_1^{\alpha} k_2^{\beta}
  \label{eq:tmunu}\\
&+&
D \left[ p_B \cdot k_1 p_{B \nu} k_{2 \mu}+  p_B \cdot k_2 p_{B \mu}  k_{1
  \nu} - p_B \cdot k_1 p_B \cdot k_2 
\frac{ k_{1 \nu} k_{2 \mu} }{k_1 \cdot  k_2} 
 -k_1 \cdot k_2 p_{B \mu} p_{B \nu} \right] \nonumber \\
&+& i C^+ \left[ k_1 \cdot
  k_2 \epsilon_{\mu \nu \alpha \beta} (k_2^{\alpha}+ k_1^{\alpha})
  p_B^\beta+
( k_{2 \mu}  \epsilon_{\nu  \alpha \beta \gamma} +k_{1 \nu} \epsilon_{\mu  \alpha
  \beta \gamma} )   k_1^{\alpha} k_2^{\beta}  p_B^\gamma\right] \nonumber \\
&+& i C^- \left[ k_1 \cdot
  k_2 \epsilon_{\mu \nu \alpha \beta} (k_2^{\alpha}- k_1^{\alpha})
  p_B^\beta+ (k_{2 \mu}  \epsilon_{\nu  \alpha \beta \gamma}-k_{1 \nu} \epsilon_{\mu  \alpha \beta \gamma})   k_1^{\alpha} k_2^{\beta}  p_B^\gamma\right] \nonumber \\
&+& i D^+ \left[ p_B \cdot k_1 \epsilon_{\mu \nu \alpha \beta}
  k_2^{\alpha} p_B^{\beta} + p_B \cdot k_2 \epsilon_{\mu \nu \alpha \beta}
  k_1^{\alpha} p_B^{\beta} +(p_{B \mu}\epsilon_{\nu  \alpha \beta
  \gamma}+p_{B \nu} \epsilon_{\mu  \alpha \beta \gamma} )  k_1^{\alpha} k_2^{\beta}  p_B^\gamma\right] \nonumber \\
&+& i D^- \left[  p_B \cdot k_1 \epsilon_{\mu \nu \alpha \beta}
  k_2^{\alpha} p_B^{\beta} - p_B \cdot k_2 \epsilon_{\mu \nu \alpha \beta}
  k_1^{\alpha} p_B^{\beta} +(p_{B \mu}\epsilon_{\nu  \alpha \beta
  \gamma}-p_{B \nu} \epsilon_{\mu  \alpha \beta \gamma} )  k_1^{\alpha} k_2^{\beta}  p_B^\gamma\right],
\nn
\end{eqnarray}
which is manifestly gauge invariant.
Here, we denote by $p_B$ the 4-momentum of the $B$-meson.
The form factors $A,B,D,C^\pm,D^\pm$ are functions of two kinematical
variables. We choose later to use the photon energies $E_1,E_2$
defined in the $B$-meson rest frame or the invariant mass $q^2$.
The amplitude squared is then given as
\begin{eqnarray}
|{\cal {A}}(B \to K \gamma \gamma)|^2 =\hspace*{-0.5cm}&&2 |A|^2-2 {\rm {Re }}(A D^*) 
~(m_B^2 k_1 \cdot k_2 -2 k_1 \cdot p_B k_2 \cdot p_B ) \label{eq:Asquared}\\
+\hspace*{-0.5cm}&& (|D|^2+2 |D^+|^2)~ (m_B^2 k_1 \cdot k_2 -2 k_1 \cdot p_B k_2 \cdot p_B )^2 
\nn\\
+\hspace*{-0.5cm}&&2 (k_1 \cdot k_2)^2 
\big| B- m_B^2 D^- + k_1 \cdot p_B(C^+ -C^-) -k_2 \cdot p_B (C^+ +C^-) \big|^2,
\nn
\end{eqnarray}
and the double differential rates (which includes a factor of $1/2$ for 
identical photons)
\begin{eqnarray}
\label{eq:E1q2}
\frac{d \Gamma}{d q^2 d E_1} =\frac{|{\cal {A}}(B \to K \gamma
  \gamma)|^2}{256 m_B^2 \pi^3}.
\end{eqnarray}
The full photon invariant mass spectrum $d \Gamma /dq^2$ can be
obtained by integration
with 
\begin{eqnarray}
\label{eq:E1bounds}
E_1^{min/max}=
\frac{1}{4 m_B} 
\bigg(m_B^2+q^2-m_K^2 \mp\sqrt{\lambda(q^2,m_B^2,m_K^2)}\bigg), 
\end{eqnarray}
where $\lambda(a,b,c)=a^2+b^2+c^2-2 ab -2 ac-2 bc$.
We also use
\begin{eqnarray}
\frac{d \Gamma}{d E_1 d E_2}& =& \frac{|{\cal {A}}(B \to K \gamma
  \gamma)|^2}{128 m_B \pi^3}. 
\end{eqnarray}

\subsection{The OPE \label{sec:ope}}

The matrix element of  $B \to K \gamma \gamma$ decays 
induced by the dipole 
operator ${\cal{O}}_7$ is a non-local one, see Fig.~\ref{fig1}, and 
difficult to estimate model-independently. 
Here our strategy is to go to 
a kinematical region where both the internal $s$- and $b$-quarks are far 
off-shell, with virtualities of order $m_b$.
To see that there exists such a region of phase space, we investigate 
the propagators of the 1PR diagrams in Eq.~(\ref{eq:W7}),
which are given as (in the bottom rest frame)
\begin{eqnarray}
(Q_1^s)^2=\hspace*{-0.5cm}&&(p^\prime+k_2)^2=m_b^2-2 m_b E_1 \, , ~~~~
(Q_1^b)^2=- \left[(p-k_1)^2-m_b^2 \right]=2 m_b E_1,   ~~~~~ \\
(Q_2^s)^2=\hspace*{-0.5cm}&&(p^\prime+k_1)^2=m_b^2-2 m_b E_2 \, , ~~~~
(Q_2^b)^2=- \left[ (p-k_2)^2-m_b^2\right] =2 m_b E_2.
\end{eqnarray}
For photon energies neither too soft nor too hard with
energies near their maximal value all $Q_i^j$ are hard.
For example, if both
photons have energies $m_b/3$, then $Q_{1,2}^s =m_b/\sqrt{3}$ and
$Q_{1,2}^b= m_b \sqrt{2/3}$.
Note that here the light quark has large energy
$m_b/3$ as well, and $q^2 =m_b^2/3$.

As a next step, we integrate out the large scales 
associated with the intermediate quark propagators.
We construct the resulting local operators out of a bottom heavy HQET quark 
$h_v$ and a strange collinear SCET quark $\chi$ \cite{Bauer:2000yr,Beneke:2002ph}.
Here,  $v=p_B/m_B$ and a light-like vector $n=p_K/E_K$, 
where $E_K, p_K$ denotes the energy, 4-momentum of the kaon.
Hence, we perform an OPE in $\Lambda_{QCD}/Q$ where
$Q=\{ m_b, E_K, Q_{1,2}^{s,b},\sqrt{q^2} \}$.
In the following we neglect $m_K$ and $m_B-m_b$.

For the lowest order 
matching onto the $b \to s \gamma \gamma$ amplitude, 
the following gauge invariant  operators with
dimension 8 with a heavy and a collinear quark and 
two photons are needed after using the
equations of motion (Wilson lines are understood in the definition of the
field $ \chi$):
\begin{align}
{\cal{Q}}_1 &= 
\frac{m_b}{4}\bar \chi \sigma_{\mu \nu}\sigma_{\alpha \beta} R h_v 
F_1^{\alpha \beta} F_2^{\mu \nu}, & {\cal{Q}}_1^\prime &= \frac{m_b}{4}\bar \chi \sigma_{\mu \nu}\sigma_{\alpha \beta} R h_v 
F_2^{\alpha \beta} F_1^{\mu \nu},\\
{\cal{Q}}_2 &=-2 i m_b \bar \chi \sigma_{\mu \nu} R h_v F_1^{\mu \nu} 
F_2^{\alpha \beta} v^\alpha n^\beta, & {\cal{Q}}_2^\prime &=-2 i m_b \bar \chi \sigma_{\mu \nu} R h_v F_2^{\mu \nu} 
F_1^{\alpha \beta} v^\alpha n^\beta,\\
{\cal{Q}}_3 &=\bar \chi \gamma^\mu L  h_v F_1^{\alpha \beta} D_\alpha \tilde F_{2 \, \beta \mu} & {\cal{Q}}_3^\prime, &=\bar \chi \gamma^\mu  L h_v F_2^{\alpha \beta} D_\alpha \tilde F_{1 \, \beta \mu},
\end{align}
where
$\tilde F_{\mu \nu} =1/2 \epsilon_{\mu \nu \alpha \beta} F^{\alpha \beta}$. 
The primed operators are related to 
the un-primed ones by interchanging photons.
Then we obtain
\begin{eqnarray}
\bar s W_7^{\mu \nu} b \epsilon(k_1)^\mu \epsilon(k_2)^\nu & \to &
-\frac{1}{2} \left \{ \left( \frac{1}{(Q_1^s)^2} -\frac{1}{(Q_1^b)^2} 
\right) {\cal{Q}}_1 + \left( \frac{1}{(Q_2^s)^2} -\frac{1}{(Q_2^b)^2} 
\right) {\cal{Q}}_1^\prime  \right.\nonumber \\ 
&+& \left.\frac{m_b E_K}{(Q_1^s)^2(Q_2^b)^2}{\cal{Q}}_2 +
\frac{m_b E_K}{(Q_2^s)^2(Q_1^b)^2}{\cal{Q}}_2^\prime  \right \}.
\end{eqnarray}
The matching of the 1PI contribution from 4-Fermi operators 
${\cal{O}}_{1,2}$ is readily
computed:
\begin{eqnarray}
\bar s W_2^{\mu \nu} b \epsilon(k_1)^\mu \epsilon(k_2)^\nu & \to &
2 \frac{ {\cal{Q}}_3 + {\cal{Q}}_3^\prime}{q^2}. 
\end{eqnarray}
Note that all  1PR and 1PI diagrams
are of the same order in $1/Q$. This is in contrast to the behavior
in $B \to \gamma \gamma$ decays, where the 1PR contribution with
intermediate $s$-quark gives the leading power result \cite{Bosch:2002bv}.

\subsection{The $B \to K \gamma \gamma$  matrix element and 
decay distribution \label{sec:spectrum}}

The matrix elements of the local operators contributing to $B \to K \gamma \gamma$ decays  are obtained from
tree level matching of the QCD onto the SCET currents 
\cite{Bauer:2000yr,Charles:1998dr,Beneke:2000wa} as
\begin{eqnarray}
\langle K(n)|\bar \chi h_v | B(v) \rangle & =& 2 E_K \zeta(E_K), \\
\langle K(n)|\bar \chi \gamma_\mu h_v |B(v) \rangle & =& 
2 E_K \zeta(E_K) n_\mu, \\
\langle K(n)|\bar \chi\sigma_{\mu \nu} h_v | B(v) \rangle & =& 
-2 i E_K \zeta(E_K) \left( v_\mu n_\nu -v_\nu n_\mu \right), \\
\langle K(n)|\bar \chi \sigma_{\mu \nu} \gamma_5 h_v | B(v) \rangle & =& 
- 2 E_K \zeta(E_K) \epsilon_{\mu \nu \alpha \beta} v^\alpha n^\beta,
\end{eqnarray}
where the form factor $\zeta$ can be identified with the QCD form factor 
in the usual parametrization, see e.g.~\cite{Ali:1999mm}, as $\zeta=f_+$.
This is the only form factor remaining in the symmetry limit, which 
enters all $B \to K$ transitions such as $B \to K \ell^+ \ell^-$ or 
$B \to K \nu \bar \nu$ decays.

We obtain then the following $B \to K \gamma \gamma$ amplitude
in terms of the parametrization given in Eq.~(\ref{eq:tmunu}) as
\begin{align}
\label{eq:A}
A&=-\kappa \zeta(E_K) Q_d m_b C_7  \frac{1}{8 E_1 E_2}
(2 E_1+2 E_2-m_B)(8 E_1 E_2- E_1 m_B-E_2 m_B),\\
B&=-\kappa 2 \zeta(E_K) \left( Q_d m_b C_7 \frac{(E_1+E_2)^2 }{8
  E_1 E_2 m_B} +(C_1 N_C+C_2) Q_u^2 \kappa_c \right), \label{eq:B} \\
D&=\kappa  \zeta(E_K) Q_d m_b C_7
\frac{(2 E_1+2 E_2-m_B)(4 E_1 E_2- E_1 m_B-E_2 m_B) }{2 E_1 E_2 (m_B-2 E_1)(m_B-2 E_2) m_B^2}, \\
C^+&=\kappa \zeta(E_K) Q_d m_b C_7  \frac{E_1-E_2}{4 E_1 E_2
  m_B^2}, \\
C^-&=-\kappa 2 \zeta(E_K) \frac{(C_1 N_C+C_2) Q_u^2 \kappa_c }{(2 E_1+2 E_2
  -m_B) m_B}, \\
D^+&=\kappa \zeta(E_K) Q_d m_b C_7 \frac{(E_1-E_2)(2 E_1+2
  E_2-m_B)}{4 E_1 E_2 m_B (m_B-2 E_1)(m_B -2 E_2)},  \\
D^-&=0, \label{eq:Dminus}
\end{align}
where
\begin{eqnarray}
\kappa \equiv +i 4 \frac{G_F}{\sqrt{2} \pi} V_{tb} V_{ts}^* \alpha_{em}.
\end{eqnarray}
Here we  use $m_b \simeq m_B$ but keep explicitly a factor of $m_b$ next to
$C_7$ from the definition of the dipole operator given in Eq.~(\ref{eq:O7}).

{}From Eqs.~(\ref{eq:kappalimit}) and~(\ref{eq:Asquared}) 
it follows that the  $B \to K \gamma \gamma$ rate vanishes for 
$q^2 \to 0$. This must be so since in the limit that one of the photon energies
vanishes the double photon rate does
because the decay $B \to K \gamma$ is forbidden.
As a further consequence, phase space integration of the 
$B \to K \gamma \gamma$ amplitude squared
is IR finite, as opposed to the inclusive $B \to X_s \gamma \gamma$
decays, which need the cancellation with the virtual electromagnetic
corrections in the low photon energy limit \cite{Reina:1997my}.

We obtain for the differential decay spectrum
\begin{eqnarray}
\label{eq:E1E2}
\frac{d \Gamma}{d E_2 d E_1}=\hspace*{-0.5cm}&&\frac{\alpha_{em}^2 G_F^2 
|V_{tb} V_{ts}^* |^2}{256 m_B \pi^5} |\zeta(E_K)|^2 
\bigg\{ m_b^2 Q_d^2 |C_7|^2
\frac{(m_B-2 (E_1+E_2))^2}{ E_1^2 E_2^2}
\nonumber \\
\times \hspace*{-0.5cm} && \bigg( 24 E_1^2 E_2^2 -4 m_B E_1 E_2 (E_1+E_2) 
+m_B^2 (E_1^2+ E_2^2) \bigg)\nonumber \\
+\hspace*{-0.5cm}&&  32 m_b  Q_d Q_u^2{\rm Re}(C_7 \kappa_c^*)(C_1 N_C + C_2) m_B (m_B-E_1-E_2)(m_B-2 (E_1+E_2)) \nonumber \\
+\hspace*{-0.5cm}&&  32 Q_u^4
|\kappa_c|^2 (C_1 N_C + C_2)^2 m_B^2 (m_B-E_1-E_2)^2 
\bigg\}.
\end{eqnarray}
We recall that Eq.~(\ref{eq:E1E2}) is valid in the domain of the 
Dalitz plot with a 
fast light meson with  energy $E_K \gg \Lambda_{QCD}$ and $Q_{1,2}^{s,b}$ and
$q^2$ hard.
%
\section{Additional contributions \label{sec:addi}}
In this section we discuss contributions to $B \to K \gamma \gamma$ decays beyond Eq.(\ref{eq:E1E2}). These are if a photon gets radiated from the 
spectator quark, which is discussed in Section \ref{sec:off} and 
annihilation contributions, see Section \ref{sec:anni}.
Effects of this type can be experimentally accessed by comparing
neutral versus charged $B$ decays.
In Section \ref{sec:long} we further discuss contributions from intermediate 
hadronic states.

\subsection{Photon radiation off the spectator \label{sec:off}}

\begin{figure}[htb]
\vskip -1.5truein
\centerline{\epsfysize=10.0in
{\epsffile{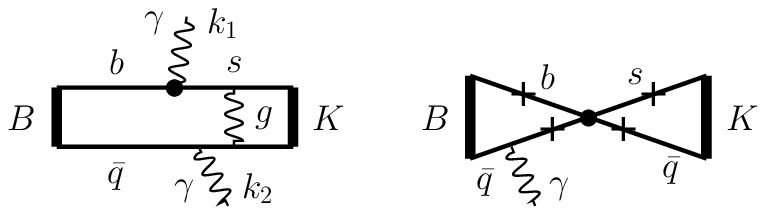}}}
\vskip -7.2truein
\caption{\it Contributions to the $B \to K \gamma \gamma$ amplitude beyond the form factor-type OPE. Left hand side: 
radiation off the spectator, the blob denotes an insertion of the 
dipole operator
${\cal{O}}_7$. Right hand side: Leading power annihilation contributions, the blob denotes a Four-Fermi operator and the second photon can be attached to any of the crosses. Diagrams with interchanged photons are not shown.}
\label{fig:OPE}
\end{figure}

The OPE is invalidated by contributions where 
one of the photons is emitted by the spectator, for an example see 
the left diagram  of Fig.~\ref{fig:OPE}.
In the case of {\it soft} gluon exchanges the momenta of the final state
$s$- and anti-quark are $\vec p_s \simeq -\vec k_1$ and
$\vec p_{\bar q} \simeq -\vec k_2$ up to order $\Lambda_{QCD}$ in the decaying heavy meson rest frame.
This configuration does not allow a Kaon to be formed if the angle 
between the photons 
is large $\sin \theta \simeq |\vec k_\perp |/E_K \gg \Lambda_{QCD}/m_B$.
This condition is indeed satisfied in the OPE region
around  $E_{1,2,K} \sim m_b/3$, where
the $B$-decay products form a Mercedes Benz star in the 
$B$-rest frame, i.e.~have angles of $120^\circ$ between each other.
Therefore, the exchange of soft gluons in these types of diagrams is 
excluded by the special kinematics.

There are further hard scattering corrections from energetic gluon 
exchange, which are, however, suppressed by the strong coupling constant. 
They can produce the $K$ also in a highly asymmetric
configuration where $p_s \simeq p_K$ and are not power suppressed.
We assume here that the endpoint suppression is sufficient to
render this contribution subdominant.
Note that due to the existence of a hard line in these diagrams, in the 
effective theory they get contributions from SCET operators 
\cite{Beneke:2003pa} involving collinear gluons.
For example, in the left diagram shown in Fig.~\ref{fig:OPE} the 
$s$-quark emitted from the FCNC vertex has virtuality 
$(Q_{2}^s)^2 \sim {\cal{O}}(m_b^2)$ and after integrating out these scales it 
is mediated by an operator with field content $\bar \chi h_v$ plus
one photon and a collinear gluon. Analogous statements hold for the other 
gluon exchange topologies which are not shown in Fig.~\ref{fig:OPE}.

\subsection{Annihilation topologies \label{sec:anni}}

If we take additional operators in the effective Hamiltonian 
Eq.~(\ref{eq:heff}) into account,
annihilations topologies in $B \to K \gamma \gamma$ decays arise.
Weak annihilation (WA) contributions in charged
$B^\pm \to K^\pm \gamma \gamma$ decays are induced by the operators
${\cal{O}}_{1,2}^u \simeq (\bar s \gamma^\mu L u)(\bar u \gamma_\mu L b)$, which
are obtained from ${\cal{O}}_{1,2}$ by replacing $c$ with $u$, and are 
CKM suppressed.
In both neutral and charged $B$ decays penguin annihilation is possible.
It is suppressed by the small Wilson coefficients of the penguin operators
which are of order $10^{-2}$ in the SM.

The leading annihilation diagrams are the ones with at least one
photon emitted from the light anti-quark in the $B$-meson, see
Fig.~\ref{fig:OPE}.
Then one propagator is off-shell by order $\sim Q \Lambda_{QCD}$.
Dimensional analysis indicates that this contribution is leading 
power with respect to the form factor-type contribution\footnote{More precisely, the scaling law of the heavy-to-light form factor 
is  $f_+ \propto m_b^{-3/2}/(1-q^2/m_b^2)^2$
\cite{Charles:1998dr}, which simplifies to the ``standard'' form at large
recoil also at $q^2 ={\cal{O}}(m_b^2)$ as long as $q^2$ is order one away 
from $m_b^2$.}
\begin{equation}
\frac{f_B f_K}{m_B \Lambda_{QCD} f_+} \propto {\cal{O}}(1) 
~~~~{\mbox{with}}~~~~
f_B \propto m_b^{-1/2}, \,~~~~~~~ f_+ \propto m_b^{-3/2},
\end{equation}
therefore it is only suppressed by small CKM elements or 
small Wilson 
coefficients. Such leading power annihilation contributions 
have also been
found in $B \to K \ell^+ \ell^-$ decays \cite{Beneke:2001at}.
However, their impact is bigger in the
two-photon decays than in the semileptonic decays since the dominant Wilson 
coefficients associated with the form factor type contribution is larger
in the latter. We estimate for $B \to K \gamma \gamma$ decays 
${\cal{A}}^{WA}/{\cal{A}}^{form factor} \sim 
V_{ub} V_{us}^*/(V_{tb} V_{ts}^*) \; C_2/C_7$, a correction of
order ten percent at the amplitude level.

\subsection{Long-distance effects \label{sec:long}}

Resonance effects through $B \to (\eta_x \to \gamma \gamma) K$ 
where $\eta_x=\eta, \eta^\prime, \eta_c$ and
$B \to (K^* \to K \gamma) \gamma$ have been discussed in
\cite{Singer:1997ti,Choudhury:2002yu}.
In particular, the latter decay chain has been used to model the 
${\cal{O}}_7$-type contribution in these works.
Relevant here to the kinematical OPE window 
are only the charmonium resonances,
since invariant di-photon masses of $m_\eta^2,m_{\eta^\prime}^2$ are too low
and the decays with virtual $K^*$ has one photon with near maximal energy
$\sim m_B/2$, which is excluded.
To be specific, the $\eta_c$ and the $\chi_{c0,c2}$ have masses in the 
signal region and sufficient branching ratios given as 
\citer{PDG2004,Aubert:2004fp} 
\begin{eqnarray}
{\cal{B}}(B \to K \eta_c) \times {\cal{B}}(\eta_c \to \gamma \gamma) \simeq 4 \cdot 10^{-7}, \\
{\cal{B}}(B \to K \chi_{c0}) \times {\cal{B}}(\chi_{c0} \to \gamma \gamma) \simeq 2 \cdot 10^{-7}, \\
{\cal{B}}(B \to K \chi_{c2}) \times {\cal{B}}(\chi_{c2} \to \gamma \gamma) <
1 \cdot 10^{-8}. 
\end{eqnarray}
We neglect in our study the $\chi_{c2}$ since the production through
$B \to K \chi_{c2}$ has not been seen yet and  its impact on
$B \to K \gamma \gamma$ is at least an order of magnitude below the other two
charmonia.

To estimate how severe the long-distance contamination is, we assume
Breit-Wigner form for the scalar and pseudoscalar 
$X=\eta_c,\chi_{c0}$ resonance amplitudes
\begin{eqnarray}
{\cal{A}}(B \to K (X \to \gamma \gamma))= 
{\cal{A}}(B \to X K)  
\frac{1}{q^2-m_{X}^2+i m_{X} \Gamma(X)} {\cal{A}}(X \to \gamma \gamma),
\end{eqnarray}
where
\begin{eqnarray}
{\cal{A}}(\eta_c \to \gamma \gamma)=i a(\eta_c) F_{\mu \nu} \tilde F^{\mu \nu}   \; ,~~~~~~~
{\cal{A}}(\chi_{c0} \to \gamma \gamma)= a(\chi_{c0}) F_{\mu \nu} F^{\mu \nu} , 
\end{eqnarray}
and take $|a(X)|$ and  $|{\cal{A}}(B \to X K)|$ from the measured rates
\begin{eqnarray}
\Gamma(X)=\frac{|{\cal{A}}(X \to \gamma \gamma)|^2 m_X^3}{16 \pi} \; ,~~~~~~~
\Gamma(B \to K X)=\frac{|{\cal{A}}(B \to X K)|^2 |\vec p_K|}{8 \pi m_B^2}.
\end{eqnarray}
This phenomenological parametrization has problems because the information 
about the strong phase is lost,
a residual $q^2$-dependence in the partial amplitudes is neglected and there
is double counting to some extend. However, the typical size of the effect is 
reproduced correctly. We use this model to 
see whether short-distance physics can be extracted from 
$B \to K \gamma \gamma$ decays in the OPE region.
This is worked out numerically in Section \ref{sec:pheno}.

The pollution from $c \bar c $ is also known from rare semileptonic decays, 
which get a sizable background through $B \to (\Psi^{(\prime ,\prime \prime,..)} \to \ell^+ \ell^-) K^{(*)}$ 
that has to be removed by kinematical cuts.
If duality would be perfectly at work, the 1PI diagram with the
charm loop would include the contributions from the charmonium resonances.
For $q^2 \simeq m_b^2 \gg m_c^2$ OPE methods have been suggested for
$B \to K^* \ell^+ \ell^-$ decays
for a model-independent calculation of the rate \cite{Grinstein:2004vb}.
For $B \to K \gamma \gamma$ decays, this amounts at lowest order
to expanding 
$\kappa(q^2,m_c^2)=\kappa(q^2,0)+{\cal{O}}(m_c^2/q^2)$, that is, using 
$\kappa_c=1/2$, see Eq.~(\ref{eq:kappalimit}). However, in the double photon
decays we cannot have $q^2$ large enough and 
hence are too close to the charm threshold to use this expansion.

\section{Phenomenology of $B \to K \gamma \gamma$ decays 
\label{sec:pheno}}

We discuss the OPE allowed  phase space and give
estimates for the  $B \to K \gamma \gamma$ branching ratio in the SM
in Section \ref{subsec:SM} and in Section \ref{subsec:NP} with NP.

\subsection{Dalitz plot and SM predictions \label{subsec:SM}} 

\begin{figure}[htb]
\vskip 0.0truein
\centerline{
\epsfysize=2.0in{\epsffile{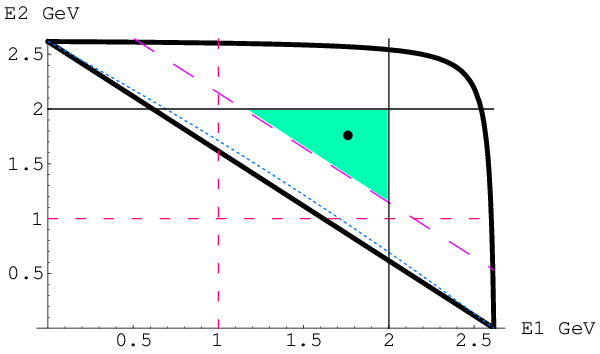}}
}
\vskip 0.0truein
\caption{\it $B \to K \gamma \gamma$ phase space in the plane of the
  photon energies $E_1$ and $E_2$ in GeV. The thick solid line is the
  phase space boundary, the thin solid, dotted, short-dashed,
  long-dashed lines denote
the cuts $ E_{1,2} <2 \, \mbox{GeV}$, $\theta >45^\circ$, $ 1 \,
  \mbox{GeV} < E_{1,2}$ and  $q^2 >m_B^2/5$.
Also shown is the point $E_1=E_2=m_B/3$.}
\label{fig:phase}
\end{figure}

The $B \to K \gamma \gamma$ Dalitz plot in the $E_1-E_2$ plane is 
shown in Fig.~\ref{fig:phase}.  The surviving
phase space after cuts is the shaded triangle. We use
$E_{1,2} < 2 \, \mbox{GeV}$ (thin solid lines) which
imply $E_K > 1.3 \, \mbox{GeV}$, and $q^2 >m_B^2/5$ (long-dashed line). 
As can be seen, other cuts required by our OPE analysis, 
such as a minimal angle
between the photon momenta for example $\theta >45^\circ$ (dotted line)
and  a low energy cut on the photon energies of 1 GeV (short-dashed
lines), are ineffective
in the presence of the previous ones. The reference point
$E_{1,2},E_K=m_B/3$ is in the surviving region.
The corresponding $q^2$-range is numerically $5.6~ \mbox{GeV}^2 < q^2 < 14.6~\mbox{GeV}^2$.
Note that the exclusion of the low di-photon mass events is not a big 
loss in the event
rate since the $B \to K \gamma \gamma$ is suppressed by small $q^2$.

\begin{figure}[htb]
\vskip 0.0truein
\begin{center}
\includegraphics[height=4.4in,width=3.3in,angle=270]{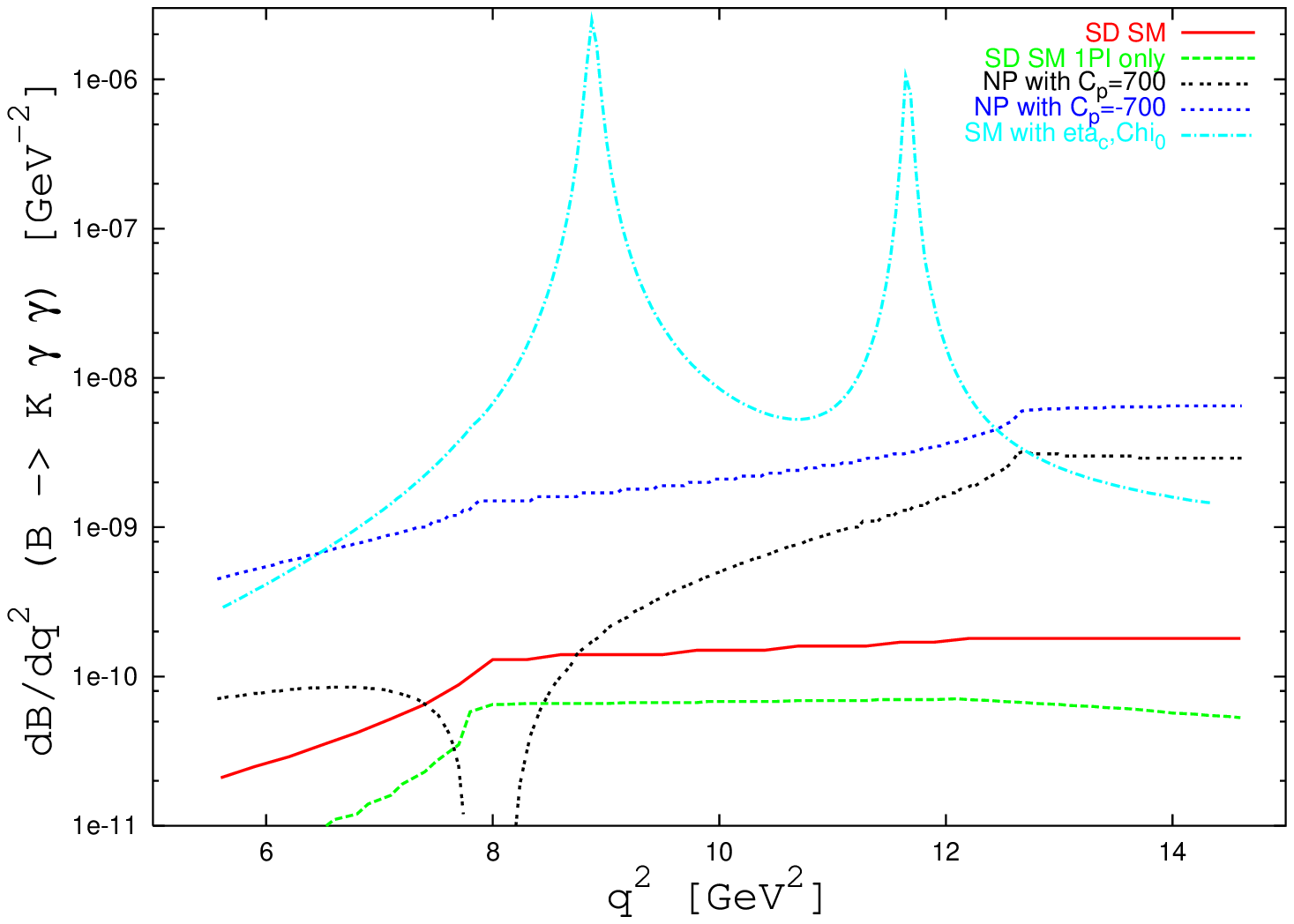}
\vskip 0.0truein
\end{center}
\caption[]{\it Di-photon mass distribution of 
$B \to K \gamma \gamma$ decays with the cuts specified in
Section \ref{sec:pheno} from the OPE. The solid (dashed) curve denote the SM pure SD contribution with (and without) the 1PR terms. The  dotted (double-dotted)
curve corresponds to a NP scenario with $ C^\tau_{P}=-700 (+700)$ and
$C^\tau_{S}=0$.
The dash-dotted curve is the SM including the resonance contributions
from the $\eta_c$ and the $\chi_{c0}$.}
\label{fig:q2}
\end{figure}

For our numerical analysis, we evaluate the SM
Wilson coefficients at leading log at $\mu=m_b$ and take the
$B \to K$ form factor $\zeta=f_+$ from Light cone-QCD sum rules
$f_+(0)=0.319$~\cite{Ali:1999mm}, which is in good agreement 
with the corresponding lattice-QCD result~\cite{Abada:1999xd}.
To be specific, we use 
$C_1=-0.25$, $C_2=1.11$ and $C_7=-0.35$.
We integrate Eq.~(\ref{eq:E1q2}) between $max(E_1^{min}, 1.2 \, \mbox{GeV})$
and $min(E_1^{max}, 2 \, \mbox{GeV})$, see Eq.~(\ref{eq:E1bounds}), to
obtain the di-photon mass spectrum  $d \Gamma/dq^2$, which is shown in 
Fig.~\ref{fig:q2}. 
As can be seen, the SD part of the spectrum (solid curve) 
is completely hidden 
behind the $\eta_c$ and the $\chi_{c0}$ resonance contribution 
(dash-dotted curve).
The integrated SD branching ratio in the SM with cuts is small,
$\triangle {\cal{B}}(B \to K \gamma \gamma)_{SM}^{OPE} \simeq 1 
\times 10^{-9}$. It has an uncertainty from the form factor of about 20 \%
and about 50 \% from the renormalization scale dependence, when $\mu$ is 
varied between $m_b/2$ and $2 m_b$. 
The order of magnitude of the branching ratio is consistent with the findings 
of Ref.~\cite{Ignatiev:2003qm}, where the 1PI contribution has  not been taken 
into account and with no cuts.

We also show for comparison the contribution
from 1PI SD diagrams only (dashed curve), 
which gives a  somewhat reduced branching ratio 
${\cal{B}}(B \to K \gamma \gamma)_{SM}^{OPE \, 1PI} \simeq 0.5
\times 10^{-9}$.  
Even if we integrate this contribution over a larger phase space region, the
resulting branching ratio is about
two orders of magnitude smaller than the corresponding ones $\sim$ 
few$\times 10^{-7}$ given in
\cite{Singer:1997ti} (where a very large value of the form factor is used) 
and \cite{Choudhury:2002yu}.
Such large $B \to K \gamma \gamma$ SD branching ratios 
are also in conflict with the one of the inclusive decays 
${\cal{B}}(B \to X_s \gamma \gamma) \simeq 3.7 \times 10^{-7}$ 
\cite{Reina:1997my}.

\subsection{New physics \label{subsec:NP}}

To be comparable to the resonance contributions at least in some region of 
phase space the SD contribution has to be lifted by roughly one order of 
magnitude above the SM one.
We investigate here whether there exists such type of NP that 
is at the same time not violating other data.
 
Beyond the SM, contributions to the dipole operator ${\cal{O}}_7$ are 
tightly constrained by
data on $b \to s \gamma$ decays. In particular, $|C_7|$ is fixed at the
30 percent level \cite{Hiller:2003js}.
The QCD penguin operators have small Wilson coefficients in the SM
$C_{SM}^{QCD}/C_2 \sim 10^{-2}$. Even an enhancement of a factor of 100, 
which is problematic because of data on $B \to K \pi$ and other 
rare decays  would constitute only an ${\cal{O}}(1)$ effect in 
$b \to s \gamma \gamma$.
Hence, these types of NP are inefficient to enhance the SD contribution 
in $B \to K \gamma \gamma$ decays to be above the $\eta_c,\chi_{c0}$
background.

Let us consider scalar/pseudoscalar operators with leptons $\ell=e,\mu,\tau$
\begin{equation}
{\cal{O}}_{S(P)}^\ell =\frac{\alpha_{em}}{4 \pi} 
\bar s R b \bar \ell (\gamma_5) \ell \, .
\end{equation}
They contribute to the matrix element of double radiative decays through
a lepton loop, similar to the charm loop shown in Figure \ref{fig1}. The
contribution w.r.t.~the charm loop scales roughly as
$\alpha_{em}/(4 \pi) \times C_{S(P)}^\ell/(Q_u^2 C_2)$.
The Wilson coefficients of the muonic operators ($\ell=\mu$) 
are constrained by ${\cal{B}}(B_s \to \mu^+ \mu^-)$ 
data to be order one \cite{Hiller:2003js}, hence way too small to raise the 
$B \to K \gamma \gamma$ SD contribution above the resonance contributions.
Assuming lepton universality, i.e., $C_{S(P)}^\ell \propto m_\ell$, 
which is realized if the operators are induced for example 
by neutral Higgs exchange, the electron
contribution is tiny and $|C^\tau_{S(P)}| \lsim 30$, which is still not large 
enough. If one relaxes lepton universality, however, one obtains 
model-independently 
$|C^\tau_{S(P)}| \lsim 700$ from
${\cal{B}}(B_s \to \tau^+\tau^-) <5 \%$ \cite{Grossman:1996qj}.
Sizable enhancements of the electron couplings are excluded by data on
$b \to s e^+ e^-$ decays.
The operators ${\cal{O}}_{S(P)}^\tau$ enter  $b \to s \gamma$ decays
as a 2-loop electromagnetic radiative correction, hence give only tiny effects
in the single photon channel.
There is no one-loop mixing of 
${\cal{O}}_{S}^\tau-{\cal{O}}_{P}^\tau$ onto ${\cal{O}}_9$,  see
\cite{Hiller:2003js} for details, hence no correction to 
$b \to s e^+ e^-$ and $b \to s \mu^+ \mu^-$ decays at this level.
Note that the 
decoupling of  $b \to s \gamma$ decays from NP in $b \to s \gamma \gamma$
has also been discussed in the context of SUSY with broken 
R-parity in $B_s \to \gamma \gamma$ and $B \to X_s \gamma \gamma$ decays
in Ref.~\cite{Gemintern:2004bw}.

Explicit calculation of the $\tau$-loop\footnote{For the OPE this requires the operators
${\cal{Q}}_4 =m_b \bar \chi R h_v F \cdot F$ and
${\cal{Q}}_5 =m_b \bar \chi R h_v F \cdot \tilde F$.
}
gives the following corrections 
to Eqs.~(\ref{eq:A})-(\ref{eq:Dminus}) as 
$A \to A+ \delta A_\tau$, $B \to B + \delta B_\tau$ etc., where
\begin{eqnarray}
\delta A_\tau&=&\kappa \frac{\alpha_{em}}{4 \pi} \zeta(E_K) 
C_S^\tau E_K m_\tau  \bigg( \kappa_\tau (4-z_\tau)+\frac{z_\tau}{2} \bigg), \\
\delta B_\tau&=&\kappa \frac{\alpha_{em}}{4 \pi} \zeta(E_K) C_P^\tau 
\frac{E_K}{m_\tau}(1-2 \kappa_\tau), 
\end{eqnarray}
and all other $\delta C^\pm_\tau$, $\delta D_\tau$, $\delta D^\pm_\tau$ vanish.
Note that the matrix elements of ${\cal{O}}_{S(P)}^\ell$ can be deduced 
from the one of the fierzed QCD penguin ${\cal{O}}_6$,
see e.g.~\cite{Chang:1997fs,Reina:1997my}.
 The impact of $C_P^\tau= \pm 700$, $C_S^\tau=0$  
and all other Wilson coefficients assuming SM values
is shown in Fig.~\ref{fig:q2} (dotted and 
double-dotted curves). As can be seen, even maximal NP enhancement  brings 
the SD distribution barely
above the resonance background outside the $\eta_c$ and $\chi_{c0}$ peaks.
For $C_P^\tau=-700$, $C_S^\tau=0$ we obtain
$\triangle {\cal{B}}(B \to K \gamma \gamma)_{NP}^{OPE} \simeq (1-2) 
\times 10^{-8}$, an order of magnitude above the SM value.
\section{Conclusions \label{sec:conclusions}}
We presented predictions for the exclusive $B \to K \gamma \gamma$ decay.
In particular, we calculated the SD contribution to the
matrix element by means of an 
OPE in $1/Q$, where $Q$ is a combined scale of order $m_b$ and 
contains the $b$-quark mass, the energy of the final meson in the $B$ 
rest frame, the photon 
invariant mass and the virtualities of the intermediate quarks in 
the 1PR $b \to s \gamma \gamma$ diagrams. Then, around the Dalitz region where 
all decay products have energy $\sim m_b/3$, the 
$b \to s \gamma \gamma$ amplitude can be expressed 
at lowest order in terms of local dimension 8 operators made out of a heavy 
HQET and a collinear SCET field plus two photons.
The resulting local matrix element is given in terms of a
form factor known from $B \to K \ell^+ \ell^-$ decays.

We found that the resulting SD branching ratio in the SM is small, 
order $10^{-9}$, which is at variance with
previous estimates \cite{Singer:1997ti,Choudhury:2002yu}.
Such small branching ratios for double radiative decays 
are not a complete surprise.
In fact, being of the same order in $\alpha_{em}$ as semileptonic rare decays
with branching ratios of few$\times 10^{-7}$ \cite{Ali:1999mm},
we expect 
\begin{eqnarray}
{\cal{B}}(B \to K \gamma \gamma) \sim \left[ 
\frac{|\kappa_c Q_u^2 C_2|^2}{|C_9|^2+|C_{10}|^2} ~~ \mbox{or} ~~ 
\frac{|C_7|^2}{|C_9|^2+|C_{10}|^2} \right] \times
{\cal{B}}(B \to K \ell^+ \ell^-) \simeq {\cal{O}}
(10^{-9}).\nn
\end{eqnarray} 
The double photon decays are
substantially suppressed with respect to the semileptonic ones, since
the di-lepton operators with
large coefficients $|C_{9,10}^{SM}| \simeq {\cal{O}}(4-5)$
are not contributing to the former, see e.g.~\cite{Buchalla:1995vs}.

In our analysis we further took long-distance effects via
$B \to \eta_c  K \to \gamma \gamma K$ and
$B \to  \chi_{c0} K \to \gamma \gamma K$
into account and found that
the SM SD contribution is not accessible behind the resonance
peaks, see Fig.~\ref{fig:q2}.
We also discussed additional contributions from photon radiation off the 
spectator and annihilation diagrams, which induce sizable theoretical
uncertainties and exhibit increased complexity compared to the ones in
$B \to K^* \gamma$ decays due to the second (energetic) photon.
On the other hand, non-standard scalar/pseudoscalar couplings to taus can give 
enhanced $B \to K \gamma \gamma$ branching ratios of the order $10^{-8}$. 
Note that such a NP scenario is consistent with data on other rare decays 
such as $b \to s \gamma$ and $b \to s \ell^+\ell^-$.
In particular, large couplings are allowed because FCNC decays into a 
tau pair are  essentially unconstrained to date.
In other words, decays such as $B_s \to \tau^+ \tau^-$ and 
$B \to K^{(*)}  \tau^+ \tau^-$ have sizable room for NP.
The corresponding maximal SD $B \to K \gamma  \gamma$ spectrum with NP 
in the $\tau$ couplings 
leaks marginally outside the resonance background. 
We conclude that it is very difficult to
test SD physics with exclusive $B \to K \gamma \gamma$ decays.

Very similar arguments apply to $B \to K^* \gamma \gamma$ decays. We
obtained for the pure SD SM estimate of the 1PI induced contribution 
${\cal{B}}(B \to K^* \gamma \gamma) \simeq \mbox{few} \times 10^{-9}$,
which is two orders of magnitude below the corresponding value quoted in 
\cite{Choudhury:2003rc}.
%

{\bf Note added:} By means of the Schouten identity \cite{schouten}
only four of the form factors in Eq.~(\ref{eq:tmunu}) correspond to
independent Lorentz structures. In fact, $C^\pm$ and $D^-$ can be absorbed
into $B$, such that redefining
$B \to B -m_B^2 D^- +k_1 \cdot p_B ( C^+ -C^-) -k_2 \cdot p_B ( C^+ +C^-)$
with the remaining form factors $A,D,D^+$ left unchanged gives identical
results for the
$B \to K \gamma \gamma$ amplitude.
%
\subsection*{Acknowledgments}
We would like to thank Thomas Becher, Gerhard Buchalla, Thorsten Feldmann, 
Olivier P\`ene and 
Amarjit Soni for discussions.
G.H.~gratefully acknowledges the SLAC theory group for hospitality
and Pierre Bleile for recovery of files from a defective hard drive.
The research of A.S.S. is supported by the Deutsche 
For\-schungs\-ge\-mein\-schaft (DFG) under contract BU 1391/1-2.
%

\end{document}